PREPRINT

# Arrhythmia Classification Using Graph Neural Networks Based on Correlation Matrix

Seungwoo Han
*Dept. of Electrical Engineering and Computer Science*
*Tokyo University of Agriculture and Technology*
seungwoohan0108@gmail.com

***Abstract*— With the advancements in graph neural network, there has been increasing interest in applying this network to ECG signal analysis. In this study, we generated an adjacency matrix using correlation matrix of extracted features and applied a graph neural network to classify arrhythmias. The proposed model was compared with existing approaches from the literature. The results demonstrated that precision and recall for all arrhythmia classes exceeded 50%, suggesting that this method can be considered an approach for arrhythmia classification.**

## I. INTRODUCTION

Electrocardiogram (ECG) is a non-invasive tool that measures the heart's electrical activity [1]. It has been used in various applications, such as arrhythmia detection, emotion recognition, and biometric identification, by leveraging machine learning and deep learning techniques [2]–[8].

Additionally, with the recent advancements in graph neural network (GNN), there has been growing research on applying GNN for data mining and representation learning. GNN is neural networks designed to learn from data represented in graph structures. A graph consists of nodes and edges, and GNN learn the relationships between these nodes and edges. GNN has been used for arrhythmia classification based on a connectivity graph structure reflecting the temporal position of ECG [9] and lead based graph classification of arrhythmias [10]. GNN offer high interpretability and performance by transforming the inherent knowledge of the data into adjacency matrices, reflecting the connectivity between features. This study proposes GNN-linear layer fusion based arrhythmia classification method that utilizes Pearson correlation matrices [11] derived from the features of ECG. In this method, features with high correlations are considered. After generating the adjacency matrix, graph features are extracted using GNN, and these are combined with features extracted from a linear neural network to classify arrhythmias.

The contribution of this paper is to create explainable artificial intelligence model for arrhythmia classification through grounded graph generation. The structure of this paper is as follows: Section 2 describes dataset collection method, feature extraction, graph generation, and the proposed model. Section 3 compares the results of previous studies with those of the proposed model. Section 4 concludes with analysis results and future research directions.

## II. MATERIAL AND METHODS

### A. Data collection

The dataset used was the MIT-BIH Arrhythmia Database from PhysioNet [12]. This database consists of 48 records collected from 47 subjects, with each record containing 30 minutes of 2-lead ECG sampled at 360 Hz, including R peak information. In this paper, in accordance with the recommendations of the Association for the Advancement of Medical Instrumentation (AAMI), four records were excluded [13]. The remaining records were divided into a training set comprising 22 records and the rest into a test set. This is summarized in Table 1.

TABLE I. TRAINING AND TESTING SET

| *Training records* | *Testing records* |
|---|---|
| 101, 106, 108, 109, 112, 114, 115, 116, 118, 119, 122, 124, 201, 203, 205, 207, 208, 209, 215, 220, 223, 230 | 100, 103, 105, 111, 113, 117, 121, 123, 200, 202, 210, 212, 213, 214, 219, 221, 222, 228, 231, 232, 233, 234 |

Furthermore, we utilized only the major classes non-ectopic beats (N), supraventricular ectopic beats (S), and ventricular ectopic beats (V). We followed the guidelines [13] for arrhythmia belonging to that class. Also, we focused solely on a lead II.

### B. Feature extraction and graph generation

We applied a moving average filter with a window of 72 data points to remove baseline noise from the ECG. This was followed by another moving average filter with a 216 data points. To further eliminate high-frequency noise, we applied a 12th-order finite impulse response (FIR) filter, passing frequencies between 0.5 Hz and 35 Hz. After filtering, we segmented the ECG signals by extracting 86 samples before and 130 samples after each R peak. Subsequently, we applied the Pan-Tompkins algorithm [14] to extract PQRST segments. A total of 20 features were then extracted for each record. The description of the features is provided in Table 2. As a result, we collected training dataset with dimensions of (50,557, 20) and a test dataset with dimensions of (49,273, 20) samples. Subsequently, we calculated the Pearson correlation coefficients for the features. If the correlation coefficient was 0.9 or higher, we set it to 1 (indicating a connection), and for values below 0.9, we set it to 0 (indicating no connection).

This process resulted in the adjacency matrix. Based on the connectivity in this adjacency matrix, we set edges between nodes, with the node features being represented by the extracted features.

### C. *Proposed model*

Our proposed model and output shape in each layer are illustrated in Figure 1.

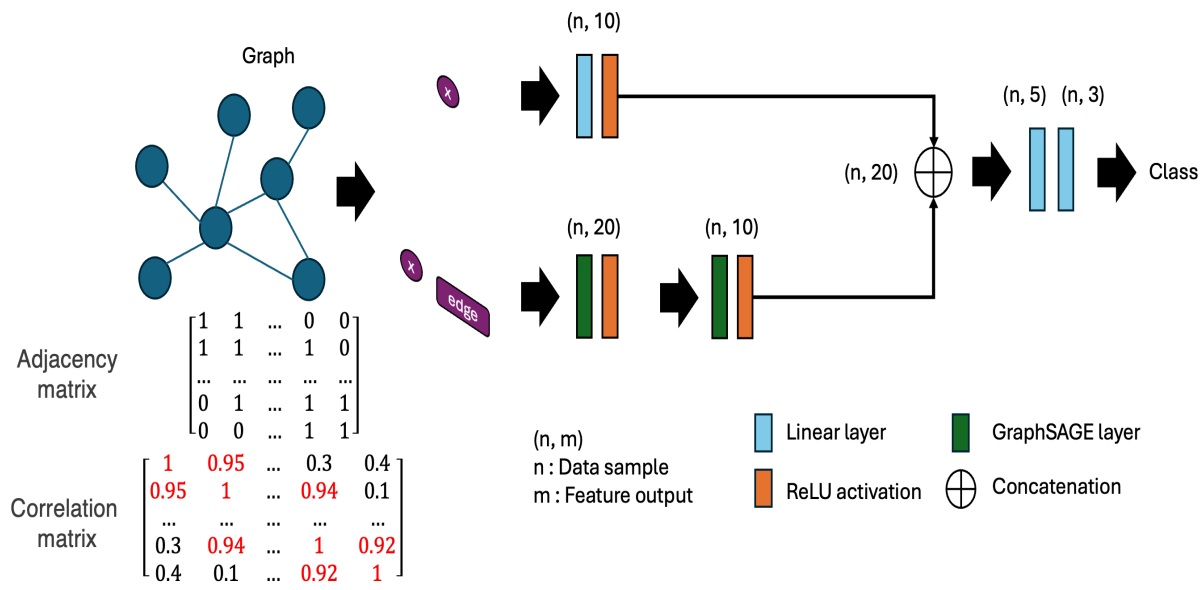

**Figure 1. Model Architecture**

We obtained feature *x* and the *edge index* from the graph and input them into both a linear and a graph neural layer. Afterward, the outputs from both networks were concatenated and passed through linear layer. For the graph neural layer, we used the GraphSAGE [15] . And we applied the rectified linear unit (ReLU) activation function [16] and adopted the Adam optimizer [17] with learning rate of 0.01, cross-entropy loss [18] and 700 epochs for model training.

TABLE II. FEATURE LISTS

| P-R amplitude | P-R time span | S-T time span | *Previous R-Present R interval |
|---|---|---|---|
| Q-R amplitude | Q-R time span | P-Q time span | *Present R-Post R interval |
| R-S amplitude | R-S time span | P-T time span | *Mean of R-R intervals |
| R-T amplitude | R-T time span | Q-T time span | *Median of R-R intervals |
| Max of each beat | Min of each beat | Variance of each beat | Root mean square of each beat |

*Work before ECG segmentation

## III. RESULT

We compared precision, recall with previous studies. The results are provided in Table 3.

TABLE III. PERFORMANCE RESULTS

| ***Methods*** | ***Precision (%)*** | | | ***Recall (%)*** | | |
|---|---|---|---|---|---|---|
| | ***N*** | ***S*** | ***V*** | ***N*** | ***S*** | ***V*** |
| Lin *et al.* [4] | 99.3 | 31.6 | 73.7 | 91.6 | 81.4 | 86.2 |
| Garica *et al.* [5] | 98.0 | 53.0 | 59.4 | 94.0 | 62.0 | 87.3 |
| Dias *et al.* [7] | **99.4** | 39.9 | **94.6** | 94.5 | **92.5** | 88.6 |
| Zhou *et al.* [8] | 98.8 | 53.8 | 92.3 | **96.9** | 89.3 | **93.3** |
| This work | 97.35 | **68.83** | 60.69 | 94.98 | 54.25 | 88.29 |

**We corrected a typo in Version [v3] of this arXiv preprint (posted on Wed, 13 Nov 2024 05:11:05 UTC) that was present in the previous version. Please note that the same typo appears in the published IEEE conference paper.

As the result, proposed model performs quite well in the N and V classes, particularly with a high precision of 97.35% and recall of 94.98% for the N class, which is the high[**] recall among the methods. In the S class, the proposed model shows the highest precision compared to other methods, but recall is lower, indicating that while the model can predict S class more accurately, there is still room for improvement in detecting all relevant instances.

## IV. CONCLUSION

This study designed a GNN-linear layer fusion model based on the correlation coefficient for arrhythmia classification. Our results showed that our model achieved a precision and recall of over 50% across all classes. This suggests the potential of our model to minimize false positives. This approach demonstrates the potential for an explainable model by generating graphs based on high correlation coefficients. Future research will focus on evaluating performance in multi-lead cases to ultimately identify the optimal graph structure.